# Kinetic analysis of phase transformations during continuous heating: Crystallization of glass-forming liquids

O.S. Houghton*

*Materials Research Laboratory, Massachusetts Institute of Technology, 77 Massachusetts Avenue, Cambridge, Massachusetts, 02139, USA*

*Corresponding author address: oshough@mit.edu

**Abstract**

Phase transformations are widely studied using continuous-heating experiments. In isothermal studies, their kinetics are often described using the Johnson-Mehl-Avrami-Kolmogorov (JMAK) rate equation. For continuous-heating studies, the same analysis has only been applied numerically. Here, a JMAK rate equation for phase transformations during continuous heating is derived. The equation is applied to the crystallization of glass-forming liquids with different kinetic behaviors and validated by comparison to experimental data and numerical simulations for the crystallization of glassy $Fe_{80}B_{20}$ (at.%). Kissinger's method of analyzing crystallization kinetics is subsequently justified, and it is shown that the non-Arrhenius temperature dependence of crystal growth rates in glass-forming liquids can be better determined by using the present model to fit the peak position, shape and height for a series of crystallization exotherms. The implications of these analytical expressions for the design and development of glass-forming systems for a broad range of applications are considered, and the application of this JMAK rate equation to other transformations during continuous heating is explored.

# 1. Introduction

Metallic, organic, inorganic and hybrid glasses have been objects of much scientific and technological interest [1, 2]. Understanding crystallization kinetics in such systems is critical to the development of high-performance glassy materials. To form a glass upon slow cooling, necessary to form thick glassy sections by casting, crystal nucleation or crystal growth rates must be low [3,4]. Though it is now possible to bypass this constraint on size by additive manufacturing or thermoplastic forming [5], heating and cooling of the glassy state after its formation risks crystallization. The need to determine crystal growth rates extends beyond processing to the technological function of certain glasses — for phase-change applications, fast switching between the crystalline and glassy phases requires rapid crystal growth during continuous heating while maintaining the ability to form a glass on cooling [6,7].

Phase transformations such as crystallization in glass-forming systems have been widely studied using continuous-heating experiments. Compared to isothermal experiments, continuous heating studies are fast, widen the temperature range in which crystallization can be measured, enable the study of fast transformations, and often mimic production processes or thermal treatments in their proposed application [6,8,9].

Although continuous-heating studies are easily performed across a wide range of heating rates, analysis of this data has required numerical simulations to fit the shape and position of crystallization exotherms and extract intrinsic material properties based on the isothermal Johnson-Mehl-Avrami-Kolmogorov (JMAK) rate equation derived over 80 years ago [7,10–12]. Numerical approaches are powerful in their ability to model complex scenarios, but analytical equations enhance the description of physical behavior and form the basis for parametric modelling. Models of this kind are computationally efficient, readily fitted to experimental data, and can form the basis for constraints in the design of high-performance materials [13]. Despite the prominence of continuous-heating experiments in the study of phase transformations, an analytical equation to describe these transformations on continuous heating has not been reported.

## 2. A JMAK equation for phase transformations during continuous heating

For an isothermal phase transformation, the JMAK rate equation describes its evolution with time assuming that:

1. there is spatially random nucleation,
2. the growth rate of the new phase is dependent only on temperature and not on time,
3. the transformation rate d$X$/d$T$ does not depend on thermal history, but only the fraction transformed $X$ and $T$,
4. the growth rate of the new phase is not limited by the removal of heat from the interface [14,15], such that the temperature of the interface does not substantially deviate from the sample as a whole.

The transformed fraction $X_{iso}$ at time $t$ during an isothermal hold at a given temperature is then given by [8,16,17]:

$$X_{\text{ISO}}(t) = 1 - \exp(-(Kt)^n), \tag{1}$$

where $K$ is a temperature-dependent kinetic coefficient, and $n$ is the Avrami exponent. The form of $K$ and the value of $n$ depends on the nature of the phase transformation. For crystallization, the exponent $n$ typically varies between 1 and 4 depending on whether: crystallization is on a fixed number of randomly dispersed nuclei or there is ongoing generation of nuclei at a fixed or decreasing rate, the nature of nucleation sites, and on whether crystal growth is one-, two- or three-dimensional [17].

During continuous-heating experiments performed using calorimetry, as the sample transforms, heat is evolved which can be measured as an exotherm on a thermogram. At any instant, the height of the peak is proportional to d$H$/d$t$, which can be assumed proportional to d$X$/d$t$ [18].

By considering a continuous-heating scan as a cycle of a small instantaneous temperature rise followed by a short isothermal hold, then the fraction transformed $X$ at a given temperature $T$ is given by:

$$\int_0^X \frac{\mathrm{d}x}{(1-x)[\ln(1-x)]^{\frac{n-1}{n}}} = \frac{n}{\Phi} \int_0^T K(T')\,\mathrm{d}T', \tag{2}$$

where it is assumed that there is no pre-existing transformed fraction, and the sample is heated from 0 K to $T$ at a constant heating rate $\Phi$.

Applying Eq. 2 to continuous heating studies requires that $K(T)$ shows a single, smooth temperature dependence [8]. The case where a sample contains a fixed number of nuclei which are randomly dispersed, and that 3D spherical growth occurs on them at a constant rate with respect to time ($n = 3$) is specifically considered. Under these assumptions, the kinetic coefficient takes the form:

$$K(T) = \left(\frac{4}{3}\pi N\right)^{\frac{1}{3}} U(T), \qquad (3)$$

where $N$ is the effective number of nuclei per unit volume and $U(T)$ is the temperature-dependent growth rate. For crystallization of a glass, where continuous normal growth occurs and the rate is limited by mobility at the crystal-liquid interface [14,15], then $U$ is given by [16]:

$$U(T) = \frac{fk_B T}{3\pi a_0^2 \eta}\left[1 - \exp\left(\frac{-\Delta G}{RT}\right)\right], \qquad (4)$$

where $\Delta G$ is the free energy difference between the liquid and crystal, $a_0$ is the average atomic diameter, $\eta$ is the viscosity of the liquid, and $f$ is the fraction of active sites at the crystal-liquid interface; $R$ and $k_B$ have their usual meanings.

At low $\Phi$, where crystallization occurs at lower temperatures and over a narrow temperature range, $\Delta G$ is large such that the temperature dependence is dominated by the viscosity. Assuming the viscosity of the liquid is well-described by the Vogel-Fulcher-Tammann (VFT) model [19–21] in the temperature region of interest:

$$(-\ln(1-X))^{\frac{1}{3}} = \frac{1}{\Phi}\left(\frac{4\pi}{3}N\right)^{\frac{1}{3}}\left(\frac{fk_B}{3\pi a_0^2 \eta_\infty}\right) F(T), \qquad (5)$$

where:

$$F(T) = \int_{T_0}^{T} T' \exp\left(-\frac{D^* T_0}{T-T_0}\right) dT', \qquad (6)$$

$$F(T) = \frac{1}{2}[T^2 - T_0^2 + D^* T_0^2 - D^* T_0 T]\exp\left(-\frac{D^* T_0}{T-T_0}\right) - \frac{D^* T_0^2 (D^*-2)}{2}\text{Ei}\left(-\frac{D^* T_0}{T-T_0}\right). \qquad (7)$$

Here, $\text{Ei}\left(\frac{D^*T_0}{T-T_0}\right)$ is the exponential integral, $D^*$ and $T_0$ are fitting parameters dependent on the composition of the glass-forming liquid, and $\eta_\infty$ is the high-temperature limit of viscosity. Using the results of Abramowitz and Stegun [22], the general result is implicitly obtained:

$$X(T,\Phi) = 1 - \exp\left[-\left(\frac{\alpha KT}{2\Phi}\right)^n\right], \tag{8}$$

where:

$$\alpha = \left[1 - D^*\frac{T_0}{T} + (D^*-1)\left(\frac{T_0}{T}\right)^2 + \frac{(D^*-2)T_0(T-T_0)}{T^2} \times \frac{\left(\frac{D^*T_0}{T-T_0}\right)^2 + 4.04\left(\frac{D^*T_0}{T-T_0}\right)+1.15}{\left(\frac{D^*T_0}{T-T_0}\right)^2 + 5.04\left(\frac{D^*T_0}{T-T_0}\right)+4.19}\right]. \tag{9}$$

Since $K(T) \propto U(T)$ for crystallization on a fixed population of nuclei, Eqs. 8 and 9 represent the crystallized fraction for one-, two- and three-dimensional crystal growth on a fixed population of nuclei during continuous heating. A full description of the possible values that the exponent $n$ can take for various conditions is provided in ref. [17].

## 3. Studying crystallization of a metallic glass

In metallic glasses (MGs), it is desirable to avoid crystallization on slow cooling to cast thick sections, and to have a large window between the glass-transition temperature $T_g$ and the peak crystallization temperature $T_x$ on continuous heating to ensure thermal stability and facilitate thermoplastic forming [5,23].

For the binary metallic glass $Fe_{80}B_{20}$ (at.%), Greer thoroughly assessed and numerically modelled its crystallization kinetics during continuous heating [12]. Not only is this glass-forming system of technological interest as a soft-magnetic material [1], but its crystallization behavior meets very closely the criteria for the present analysis to be valid. Over the temperature range of crystallization, Greer suggests that $U(T)$ shows Arrhenius temperature dependence [12]. For Arrhenius behavior of this kind, Eq. 6 simplifies to:

$$F(T) = \int_0^T T' \exp\left(-\frac{E_\eta}{RT}\right) dT', \tag{10}$$

where $E_\eta$ describes the activation energy for viscous flow. Solving this equation, the parameter $\alpha$ becomes:

$$\alpha = 1 - \left( \frac{\left(\frac{E_\eta}{RT}\right)^2 + 3.04\left(\frac{E_\eta}{RT}\right)}{\left(\frac{E_\eta}{RT}\right)^2 + 5.04\left(\frac{E_\eta}{RT}\right) + 4.19} \right), \tag{11}$$

which can also be obtained by taking Eq. 9 in the limit of $T \gg T_0$ for which $E_\eta = RT_0 D^* [1 - (T_0/T)]^{-1}$.

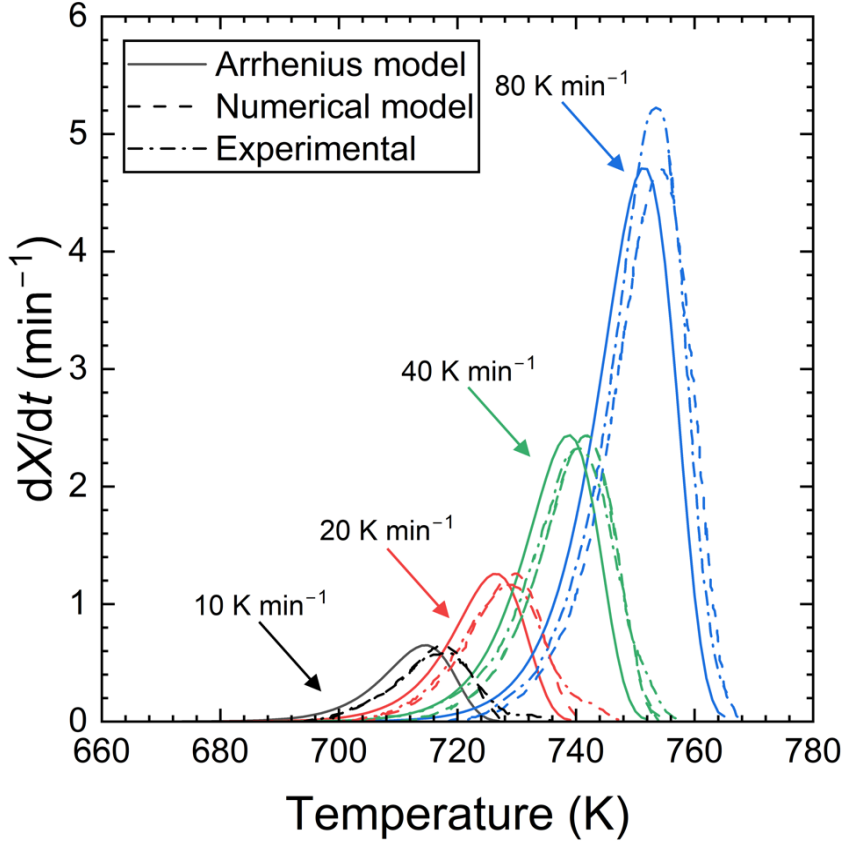

**Figure 1**. Experimental results for as-spun $Fe_{80}B_{20}$ ribbons continuously heated at rates between 10 and 80 K min$^{-1}$ reproduced from ref. [12] compared to numerical models performed in ref. [12] and the Arrhenius model described by using Eq. 8 and 11 using values for $K_0$ and $E_\eta$ reported in ref. [12].

Using the measured values for $E_\eta$ and the pre-factor for $K$ obtained by Greer, Eq. 8 can reproduce the peak shape and position for continuous-heating scans at different rates performed on as-spun $Fe_{80}B_{20}$ ribbons (Figure 1): this analytical model shows good agreement with previous experimental and numerical models [12]. In Figure 1, Eq. 8 underestimates the experimentally observed $T_x$ by 4–6 K. Differences between experimental and model thermograms could be partly attributed to the boundary conditions of integration, the assumption of Arrhenius behavior and ellipsoidal growth [12]. Differences between numerical and this analytical model are primarily due to the assumed boundary conditions of integration. When the peak shape and position of the four crystallization exotherms are fitted using a VFT model, an improved fit to the experimental data is obtained compared to the numerical and analytical Arrhenius models (Figure 2a). $Fe_{80}B_{20}$ has a finite value of $T_0$ which suggests deviations from Arrhenius behavior are observed, and are detectable over the width of a crystallization exotherm when the peak shape is fitted using the present analysis.

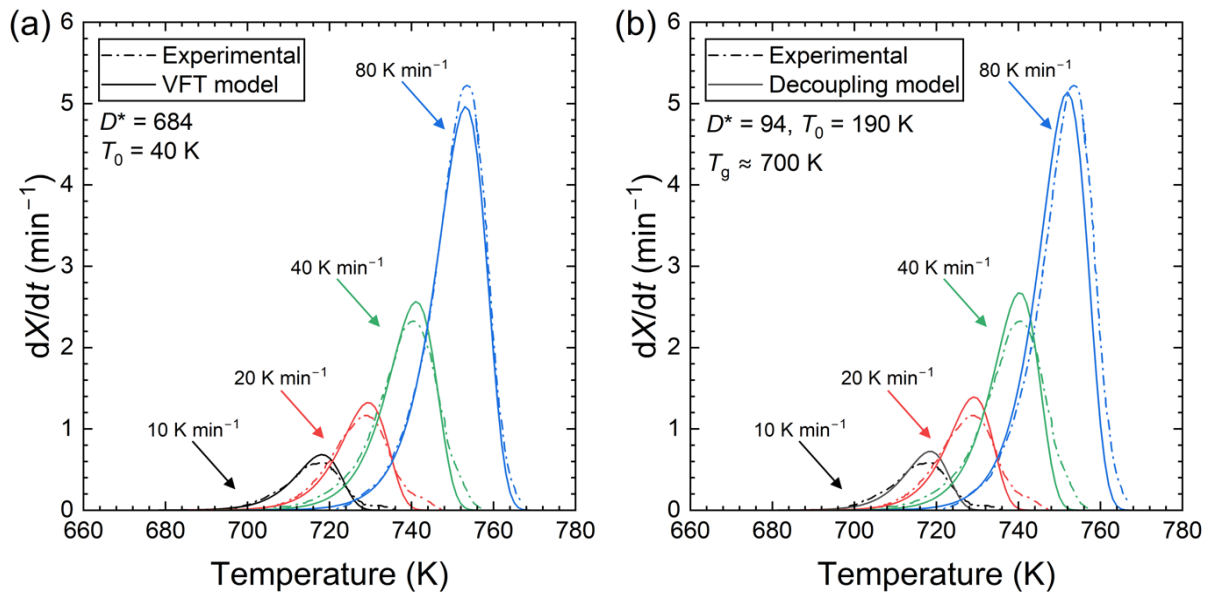

**Figure 2.** (a) Experimental results for as-spun $Fe_{80}B_{20}$ ribbons continuously heated at rates between 10 and 80 K min$^{-1}$ reproduced from ref. [12] are compared to analytical calculations performed using Eqs. 8 and 9 with values for $K_0$ reported in ref. [12] and $D^*$ and $T_0$ values of 684 and 40 K, respectively. This combination of $D^*$ and $T_0$ corresponds to an average $E_\eta$ of 241 kJ mol$^{-1}$ over the temperature region of interest. (b) shows comparison of experimental results with the crystal growth model accounting for viscous decoupling [24] (Eqs. 8 and 24) and the VFT model for viscosity with $D^*$ and $T_0$ values of 94 and 190 K, respectively.

## 4. Factors affecting resistance to crystallization

The present work highlights that the purpose of many numerical models has been to determine the value of $\alpha$ (Eqs. 9 and 11) by varying the temperature dependence of viscosity, the effective density of nuclei $N$ and the pre-factor for crystal growth. At heating rates used in conventional calorimetry, the correction for the effect of thermodynamics on crystal growth rates is negligible such that the heating rate dependence of crystallization is dominated by the temperature dependence of viscosity [24,25].

The temperature dependence of viscosity is described by the fragility of the liquid $m$ [26], which for VFT kinetics is given by:

$$m = \frac{\mathrm{d}\log\eta}{\mathrm{d}\left(\frac{T_g}{T}\right)} = \frac{D^* T_0}{\ln 10} \frac{T_g}{(T_g - T_0)^2}. \tag{12}$$

For low heating rates, where $T_x$ lies close to $T_g$, the effective activation energy for viscous flow $E_\eta$ increases as the liquid becomes more fragile (higher $m$). This corresponds to a lower resistance to crystallization on heating. With increasing $E_\eta$, the value of $T_x$ can only be maintained by reducing the value of $N$ (Figure 3a,b). When $N$ remains unchanged, higher $E_\eta$ (higher $m$) leads to an earlier onset of crystallization with a sharper and more intense exotherm [27].

The present analysis relies on the assumption referred to by Cahn as *site saturation* [28]; all nucleation occurs early in the reaction. In the present case, if $N$ is large, then the effect of ongoing nucleation during continuous heating is negligible such that site saturation is effectively achieved. For many glass-forming systems, a substantial population density of quenched-in nuclei is formed during cooling to form a glass [29]. Site saturation is achieved such that Eq. 8 provides a reliable prediction of peak shape and position. As $N$ increases (larger $K_0$ in ref. [12]), an earlier onset of crystallization is observed, $T_x$ decreases but the shape of the exotherm is not substantially changed (Figure 3c,d). The effect of changing $N$ on peak shape (Figure 3d) appears less significant than $m$ (Figure 3b).

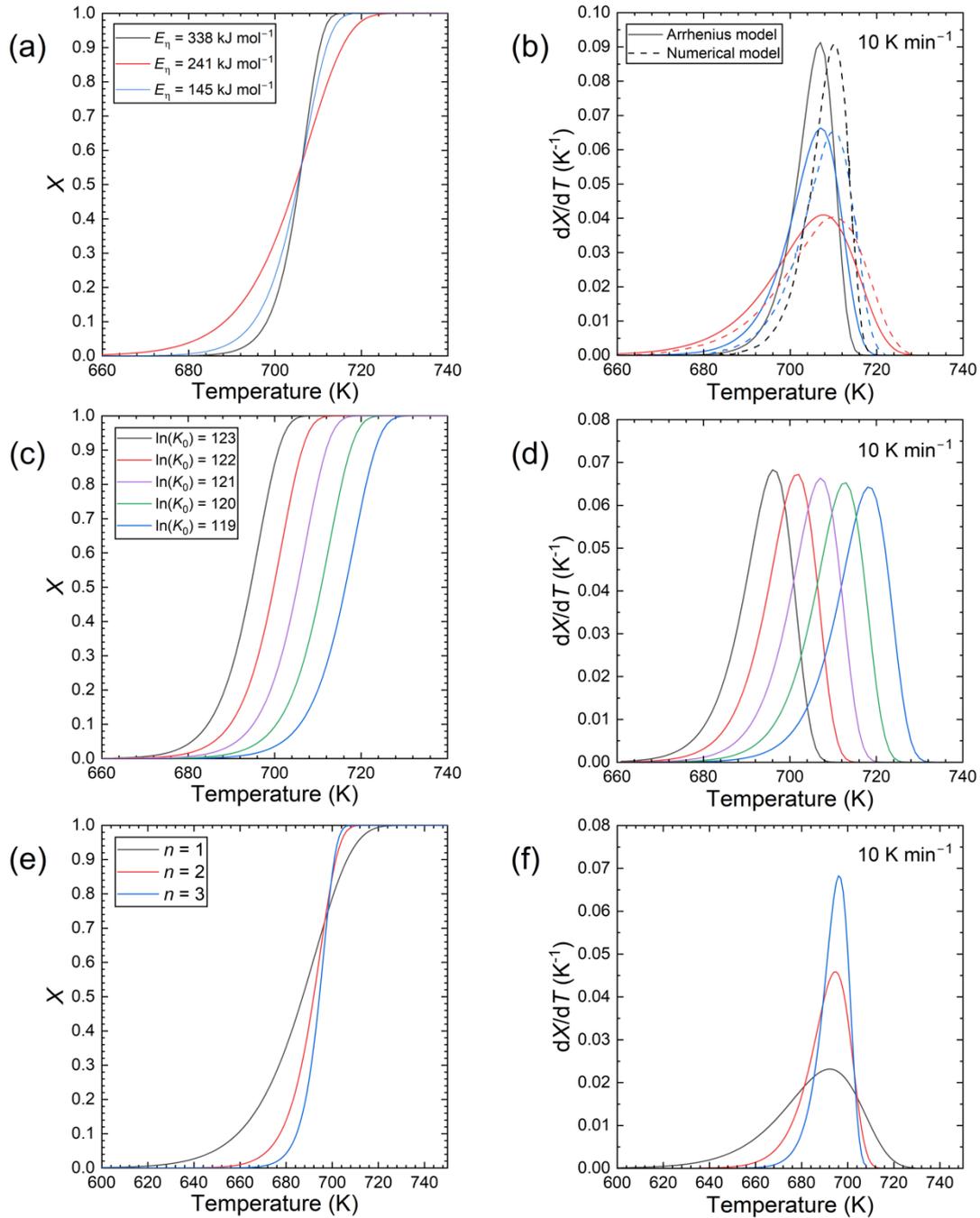

**Figure 3.** (a) Calculations of the crystallized fraction $X$ on continuous heating at 10 K min$^{-1}$ for different $E_\eta$ values and adjusting $K_0$ (proportional to $N^{1/3}$) to maintain a constant value of $T_x$. Data taken from ref. [12]. (b) Predicted shape of exotherm from (a) compared to numerical simulations performed in ref. [12]. (c) Calculations of the crystallized fraction $X$ on continuous heating at 10 K min$^{-1}$ for different $K_0$ values used in ref. [12]. The values of $K_0$ used in the fitting have been adjusted to account for differences in definition between the present work and ref. [12]. (d) Predicted shape of exotherm from (c). (e) Calculations of the crystallized fraction $X$ on continuous heating at 10 K min$^{-1}$ for fixed $E_\eta$ values and $K_0$ but different values of $n$. (f) Predicted shape of exotherm from (e).

For exceptional glass-formers, or for liquids quenched to form a glass at high cooling rates, the as-quenched glass may contain a minimal number of nuclei [29]. For such samples, site saturation is not achieved and the effects of ongoing nucleation must be considered. For the case of constant nucleation rate on which three-dimensional crystal growth occurs, the kinetic coefficient takes the form:

$$K(T) = \left(\frac{\pi N}{3}\right)^{\frac{1}{4}} I^{1/4} U^{\frac{3}{4}}, \qquad (13)$$

where $I$ is the steady-state nucleation rate. The effects of transient nucleation are assumed to be minimal as a population of embryos is inherited from quenching and prior temperature steps during continuous heating [12,30]. When the system necessitates that $K$ takes the form of Eq. 13, there is no longer a single activation energy; a precise solution of Eq. 2 is best achieved numerically.

When $N$ is kept fixed, and $K$ is adjusted only for geometric differences between 1D, 2D and 3D crystal growth, decreasing $n$ leads to a subtle shift in $T_x$ to lower temperatures. The peak becomes broader and less intense (Figure 3e,f) due to the expected differences in the geometry of crystal growth.

## 5. Kissinger method applied to crystallization

For most glass-forming liquids, a substantial population density of nuclei exists from quenching or is generated during continuous heating at low $\Phi$ before the onset of the crystallization exotherm — the effective population density of nuclei $N$ is high such that *site saturation* condition is achieved.

Kissinger's method to study crystallization as a function of heating rate has been widely adopted [31]. On this plot of $\ln(\Phi/T_x^2)$ versus $1/T_x$, the gradient is equal to $-E_x/R$ where, $T_x$ is the peak of the crystallization exotherm and $E_x$ is the effective activation energy for crystallization. Using Eq. 8, it is possible to advance the original validation of the Kissinger method for crystallization during continuous heating by Henderson [8], defining the exact form of the curve:

$$\ln\left(\frac{\Phi}{T_x^2}\right) = \ln\left(\frac{1}{2}\right) + \ln\left(\frac{\alpha K}{T_x}\right) - n\ln(-\ln[1 - X(T_x, \Phi)]). \qquad (14)$$

$T_x$ typically corresponds to a transformed fraction close to 63% (see Supplementary Material), such that the equation simplifies to the form:

$$\ln\left(\frac{\Phi}{T_x^2}\right) = \frac{1}{3}\ln\left(\frac{\pi N}{6}\right) + \ln\left(\frac{fk_B}{3\pi a_0^2}\right) + \ln\alpha - \ln(\eta), \quad (15)$$

for all $\Phi$ where $|\Delta G|$ is large. The first two terms are temperature-independent and the variation in $\alpha$ is typically small compared to the variation in $\eta$. Under these assumptions, the gradient of the Kissinger plot is then given by:

$$E_x(T_X) \approx RT_g \ln 10 \frac{d\log\eta}{d(T_g/T_x)} = mRT_g \ln 10 \left(\frac{T_x}{T_g}\right)^2 \left(\frac{T_g - T_0}{T_x - T_0}\right)^2. \quad (16)$$

In the limit of $T_0 \ll (T_g, T_x)$ and a narrow range of $\Phi$, $E_x$ does not show any temperature dependence. With accurate determination of $T_g$ and $T_0$, it is possible to determine a value of $m$ from a Kissinger plot, as implied in the work of Chen [25] and discussed by Henderson [8].

## 6. Effects of fragility and fast crystal growth

Though this method is specifically applied in this work to crystallization in metallic glass-forming systems, the method is, in principle, applicable to a phase transformation in any system where JMAK kinetics are a valid description. Glassy materials that exhibit fast crystal growth are ideal materials for phase-change applications. For these systems, which are clearly fragile ($m > 80$ [6]), Ediger *et al.* highlight that crystal growth rates are decoupled from the viscosity of the liquid [24].

In a classical approach, the crystal growth rate varies with the mobility at the crystal-liquid interface. This is assumed proportional to the atomic diffusivity in the bulk liquid, which itself is assumed to be inversely proportional to the viscosity of the liquid. This allows application of the Stokes-Einstein relation to determine an expression for $U$. At high temperatures, these assumptions generally hold. As the temperature lowers, it is suggested that progressive decoupling of atomic diffusivity from viscous flow occurs, such that the atomic diffusivity is much higher than expected from the Stokes-Einstein relation [24]. The crystal growth rate should then be given by:

$$U(T) = \frac{a}{\tau_0}\left(\frac{\eta(T)}{\eta_0}\right)^{-\epsilon} \times \exp\left[-\frac{\Delta S_f}{R}\right]\left[1 - \exp\left(\frac{-\Delta G}{RT}\right)\right], \quad (20)$$

where $a$ is the lattice spacing of the forming crystal, $\tau_0$ is the structural relaxation time at the reference viscosity $\eta_0$, $\Delta S_f$ is the entropy of fusion and $\epsilon$ is the empirical decoupling factor given by [24]:

$$\epsilon = 1.1 - 0.005m. \quad (21)$$

For many glass-forming systems, $\epsilon$ lies close to one such that its effect is not significant. For fragile systems, $\epsilon$ can become much lower than one having a substantial effect on predictions of crystal growth rate [6,24]. The VFT model for viscosity enables simple inclusion of this parameter in the present analysis. Using Eq. 20, a revised value of $\alpha$ takes the form:

$$\alpha = \frac{2(T-T_0)}{T^2}\left[1 - \frac{\left(\frac{\epsilon D^* T_0}{T-T_0}\right)^2 + 4.04\left(\frac{\epsilon D^* T_0}{T-T_0}\right) + 1.15}{\left(\frac{\epsilon D^* T_0}{T-T_0}\right)^2 + 5.04\left(\frac{\epsilon D^* T_0}{T-T_0}\right) + 4.19}\right]. \quad (24)$$

Eq. 24 and an appropriate expression for $K$ in Eq. 8 can reasonably fit the crystallization exotherms for $Fe_{80}B_{20}$ at different heating rates (Figure 2b). Numerical simulations are not required to explore the effects of decoupling from viscous flow on the shape and form of crystallization exotherms. The effects of fragility can be determined using analytical methods; parametric modelling and fittings of experimental results are possible.

## 7. Characterization of devitrification in glasses

When crystallization of a glass during continuous heating exhibits a single temperature dependence, it is shown that the Kissinger method can then be applied. The resulting plot (Figure 4a) can be related to both the population density of nuclei and the temperature dependence of the crystal growth rate. For a glass-forming liquid whose crystal growth rate is well-described by Eq. 4, values for $N$ and the effective activation energy for viscous flow $E_\eta$ can be determined from the Kissinger plot using Eq. 15.

One can consider $N$ as the population density of nucleation centers on which crystallization occurs. As such, it can represent the combined density of quenched-in nuclei, nuclei generated during annealing, or impurities that can behave as heterogeneous nucleants [33]. Assuming there is not substantial crystal growth upon these nucleation centers prior to measurement of the thermogram, the effects of impurities and thermal history on the position of $T_x$ should be readily quantifiable using the present analysis. The value of the Avrami exponent can also provide important and complementary information about the nature of nucleation sites, as outlined by Christian [17].

In fitting the peak shape and position of the crystallization exotherm for a given heating rate $\Phi$, it is possible to obtain crystal growth rates over a wider temperature range compared to the Kissinger method without the need for further experiments (Figure 4a). The shape and height of the crystallization exotherm provides important information about the non-Arrhenius temperature dependence which can be described using the VFT model in the present analysis — each combination of $D^*$ and $T_0$ produces a unique exotherm (Figure 4b).

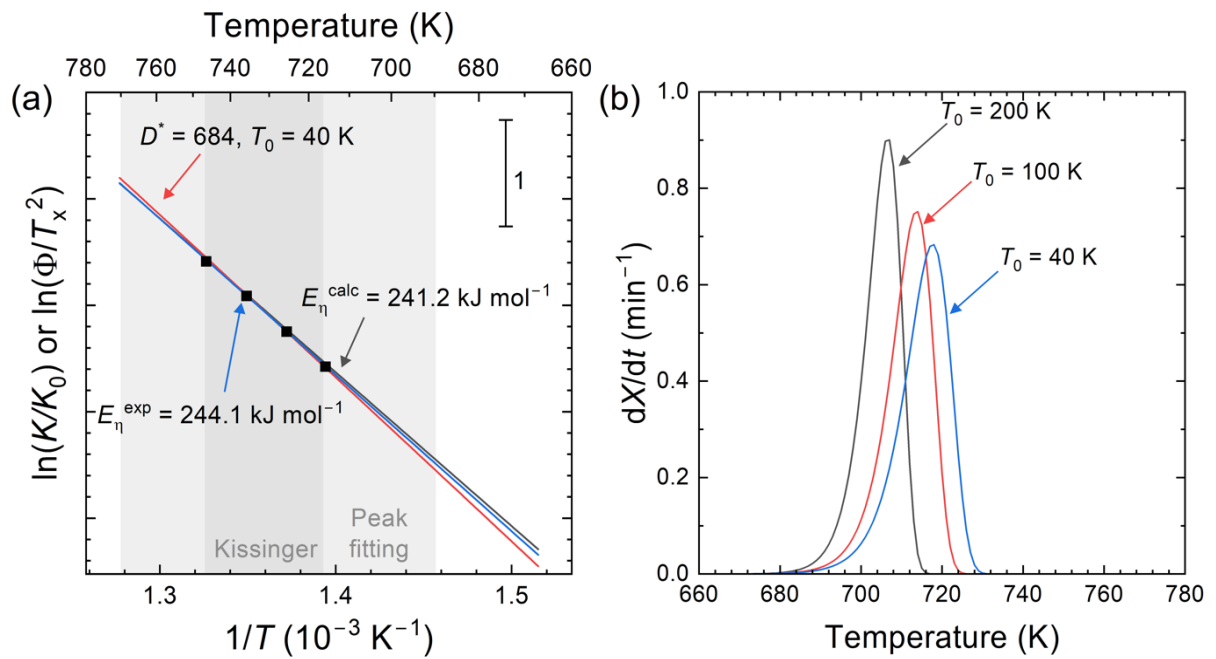

**Figure 4.** (a) Plot of $\ln(K/K_0)$ for the VFT model obtained by peak fitting in the present work, the Arrhenius model obtained by numerical fitting of the peaks and via the Kissinger method in ref. [12]. $T_x$ data plotted using the Kissinger method in ref. [12] is also shown. The temperature ranges in which data is obtained by the Kissinger method and peak fitting methods for the same series of exotherms are shaded. (b) Plot of expected crystallization exotherms on heating at 10 K min$^{-1}$ for different combinations of $D^*$ and $T_0$ that would correspond to an effective activation energy of 242 kJ mol$^{-1}$. Values of $D^*$ are 684 ($T_0 = 40$ K), 247 ($T_0 = 100$ K) and 101.4 ($T_0 = 200$ K).

For $Fe_{80}B_{20}$, eutectic crystal growth occurs [12] and so the temperature dependence of the crystal growth rate does not solely depend on mobility at the crystal-liquid interface. $D^*$ and $T_0$ parameters reported here only describe the local temperature dependence of $U$ in this region of interest. For systems whose crystal growth rate is limited by mobility at the crystal-liquid interface, deviations in $\ln(K/K_0)$ from Arrhenius behavior, which can be detected using this peak fitting approach (Figure S1), are directly related to the fragility of the liquid.

This analysis focuses on thermograms measured at low heating rates, but ultra-fast (flash) calorimetry has become a powerful technique able to achieve continuous heating and cooling at rates exceeding $10^3$ K s$^{-1}$. It enables mimicking of rapid solidification, study of degradation-prone materials, glassy states with a minimal number of quenched-in nuclei and stochastic nucleation effects [29,30,32,34]. Compared to conventional calorimetry, crystallization is studied in flash calorimetry across a wider range of heating rates, for which $E_x$ is typically not constant [6,34,35]. The Kissinger plot of $T_x$ data collected using flash calorimetry hence reflects the curvature in $\ln(K)$. The equations derived in the present work are applicable to flash calorimetry studies, but there are limitations imposed by the assumed boundary conditions. At high heating rates, crystallization occurs at high temperature and over a wider temperature range. $\Delta G$ reduces in magnitude and varies substantially within this temperature range such that thermodynamic corrections are required before attempting a Kissinger plot of the kinetic component of $U$ [6]. At even higher heating rates, the overlap of the crystallization exotherm and melting leads to incomplete crystallization, for which numerical simulations are required [6,7].

Though the primary focus of this work has been the fitting and characterization of experimental results to validate the analysis, a JMAK rate equation for continuous heating of this kind realizes opportunities to predict (with quantified uncertainty) the performance and key features of thermograms for glassy materials. Expressions for the critical cooling rate for glass formation have been established, but efforts to predict other key indicators of glass-forming ability, such as $\Delta T_x$ (equal to $T_x$–$T_g$) have been less successful [5]. The various equations within this work could enable improved prediction of crystallization behavior on continuous heating for a broad range of glass-forming systems.

## 8. Broader application to the characterization of transformations

The JMAK equation for continuous heating is potentially applicable to a broad range of phase transformations in different materials, as well as beyond materials science [36]. For phase transformations in materials, the assumptions for JMAK to be valid hold for polymorphic changes, discontinuous precipitation, eutectic and eutectoid reactions and interface-controlled growth amongst others [17]. Such transformations may be studied using calorimetry, or other techniques where transformed fractions are readily determined, such as, but not limited to dilatometry, thermogravimetric analysis and X-ray diffraction.

Isothermal heat treatments are a common post-synthesis step in the processing of high-performance materials. The advent of techniques such as additive manufacturing and flash sintering which have distinctly non-isothermal profiles can lead to distinctly different microstructures whose evolution cannot be described using a simple isothermal model. It is worthwhile considering the suitability of the present analysis to describe and model such processing techniques as a single scan or series of continuous heating scans. Physics-based parametric models of this kind are beneficial to the design of compositions that can produce the desired microstructures via these advanced manufacturing techniques.

## 9. Conclusions

A Johnson-Mehl-Avrami-Kolmogorov (JMAK) rate equation is derived to describe phase transformations during continuous heating. Applicable to a broad range of phase transformations, explicit expressions are derived for crystallization in glass-forming systems that show different kinetic behaviors. The expressions are shown to successfully model the crystallization behavior of the metallic glass $Fe_{80}B_{20}$ (at.%) and can fully justify the Kissinger method for studying crystal growth during continuous heating. These analytical expressions can be applied to understand and model the effects of kinetic fragility of the liquid, and enhanced mobility at low temperatures on the crystallization behavior of glass-forming liquids. By fitting the peak shape for a series of crystallization exotherms measured at different heating rates using this parametric model, it is possible to efficiently and more accurately determine the non-Arrhenius temperature dependence of crystal growth rates for a glass-forming liquid than by the Kissinger method alone. The derived expressions form an important link between material properties and phase transformation kinetics that could help facilitate fast and easy characterization and reliable prediction of performance and processability of glassy materials.

It is highlighted that similar expressions could be derived based on this model to characterize and predict the kinetics of other transformations which occur in a broad range of material systems.

## Acknowledgements

It is a pleasure to thank Prof. A. Lindsay Greer for many useful discussions, critical reading of the manuscript and thoughtful suggestions, and Prof. Gregory B. Olson for allowing time to pursue this work.

## Author Declarations

The author has no conflicts to disclose.

## Data Availability

The data that support the findings of this work are available from the author upon reasonable request.

# Kinetic analysis of phase transformations during continuous heating: Crystallization of glass-forming liquids
## Supplementary Material

O.S. Houghton

The Kissinger plot of $\ln(\Phi/T_x^2)$ against $1/T_x$ requires accurate determination of $T_x$ in terms of $dX/dT$. According to Borchardt [15], the peak of the crystallization exotherm at $T_x$ corresponds to the maximum transformation rate:

$$\left.\frac{d^2 X}{dt^2}\right|_{T=T_x} = 0. \tag{S1}$$

Using Eq. 8, we find:

$$\left(\frac{dX}{dt}\right) = \Phi \frac{dX}{dT} = n\Phi[-\ln(1-X)][1-X]\left[\frac{d\ln\left(\frac{\alpha KT}{2\Phi}\right)}{dT}\right], \tag{S2}$$

$$\left(\frac{dX}{dt}\right) = \Phi \frac{dX}{dT} = n\Phi[-\ln(1-X)][1-X]\left[\frac{d\ln(\alpha)}{dT} + \frac{d\ln(U)}{dT} + \frac{1}{T}\right]. \tag{S3}$$

Calculating the second derivative yields:

$$\left(\frac{d^2 X}{dt^2}\right) = \Phi\frac{d}{dT}\left(\Phi\frac{dX}{dT}\right) = \Phi^2 \frac{d^2 X}{dT^2} = n\Phi^2 \frac{d}{dT}\left[\left(\frac{\alpha KT}{2\Phi}\right)^n \exp\left[-\left(\frac{\alpha KT}{2\Phi}\right)^n\right]\left[\frac{d\ln(\alpha)}{dT} + \frac{d\ln(U)}{dT} + \frac{1}{T}\right]\right], \tag{S4}$$

$$\left.\frac{d^2 X}{dt^2}\right|_{X=X_p} = n\Phi^2 \left(\frac{\alpha KT}{2\Phi}\right)^n \exp\left[-\left(\frac{\alpha KT}{2\Phi}\right)^n\right]\frac{d}{dT}\left[\left[\frac{d\ln(\alpha)}{dT} + \frac{d\ln(U)}{dT} + \frac{1}{T}\right]\right] + n\Phi^2\left(\frac{\alpha KT}{2\Phi}\right)^n\left[\frac{d\ln(\alpha)}{dT} + \frac{d\ln(U)}{dT} + \frac{1}{T}\right]\frac{d}{dT}\left[\exp\left[-\left(\frac{\alpha KT}{2\Phi}\right)^n\right]\right] + n\Phi^2\left[\frac{d\ln(\alpha)}{dT} + \frac{d\ln(U)}{dT} + \frac{1}{T}\right]\exp\left[-\left(\frac{\alpha KT}{2\Phi}\right)^n\right]\frac{d}{dT}\left[\left(\frac{\alpha KT}{2\Phi}\right)^n\right] = 0, \tag{S5}$$

$$\left.\frac{d^2 X}{dt^2}\right|_{X=X_p} = -n\Phi^2 \ln(1-X_p)[1-X_p]\frac{d}{dT}\left[\left[\frac{d\ln(\alpha)}{dT} + \frac{d\ln(U)}{dT} + \frac{1}{T}\right]\right] - n\Phi^2 \ln(1-X_p)\left[\frac{d\ln(\alpha)}{dT} + \frac{d\ln(U)}{dT} + \frac{1}{T}\right]\frac{d}{dT}\left[\exp\left[-\left(\frac{\alpha KT}{2\Phi}\right)^n\right]\right] + n\Phi^2\left[\frac{d\ln(\alpha)}{dT} + \frac{d\ln(U)}{dT} + \frac{1}{T}\right][1-X_p]\frac{d}{dT}\left[\left(\frac{\alpha KT}{2\Phi}\right)^n\right] = 0, \tag{S6}$$

$$\ln(1-X_p)[1-X_p]\left[\left[\frac{d^2\ln(\alpha)}{dT^2} + \frac{d^2\ln(U)}{dT^2} - \frac{1}{T^2}\right]\right] - n\ln(1-X_p)^2[1-X_p]\left[\frac{d\ln(\alpha)}{dT} + \frac{d\ln(U)}{dT} + \frac{1}{T}\right]^2 - n\left[\frac{d\ln(\alpha)}{dT} + \frac{d\ln(U)}{dT} + \frac{1}{T}\right]^2 [1-X_p]\ln(1-X_p) = 0. \tag{S7}$$

Eq. S7 is satisfied when $X_P$ takes the value of zero (no crystallinity), one (fully crystallized) or:

$$X_p = 1 - \exp\left[-1 + \frac{1}{n}\left[\frac{d^2\ln(\alpha)}{dT^2} + \frac{d^2\ln(U)}{dT^2} - \frac{1}{T^2}\right]\left[\frac{d\ln(\alpha)}{dT} + \frac{d\ln(U)}{dT} + \frac{1}{T}\right]^{-2}\right]. \tag{S8}$$

For viscosity that shows near-Arrhenius behavior, and crystal growth rates not restricted by thermodynamic effects, we can take the limit that the temperature dependence of $\ln(\alpha)$ is small compared to $\ln(U)$ such that:

$$X_p \approx 1 - \exp\left[-1 + \frac{1}{n}\left[\frac{d^2\ln(U)}{dT^2} - \frac{1}{T^2}\right]\left[\frac{d\ln(U)}{dT} + \frac{1}{T}\right]^{-2}\right], \tag{S9}$$

where for large $\Delta G$ and $n = 3$:

$$\frac{d\ln(U)}{dT} = \frac{d}{dT}\left[\ln(T) - \frac{D^*T_0}{T-T_0}\right] = \frac{1}{T} + \frac{D^*T_0}{(T-T_0)^2}, \tag{S10}$$

$$\frac{d^2\ln(U)}{dT^2} = -\frac{1}{T^2} - \frac{2D^*T_0}{(T-T_0)^3}, \tag{S11}$$

$$X_p \approx 1 - \exp\left[-1 - \frac{2}{3}\left[\frac{(T_X-T_0)}{D^*T_0}\right]\right] = 1 - \exp\left[-1 - \frac{2T_g}{3m(T_X-T_0)\ln 10}\right]. \tag{S12}$$

Eq. S12 yields $X_p = 0.636$ for $Fe_{80}B_{20}$, in good agreement with ref. [8].

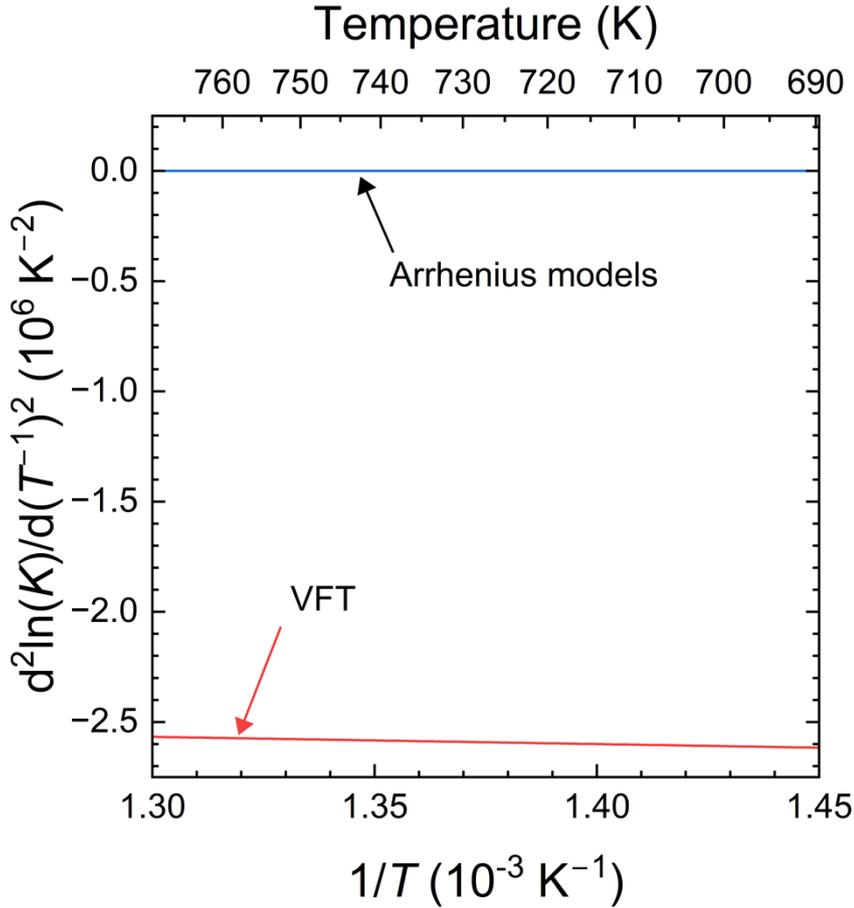

**Figure S1.** Plot of 2$^{nd}$ derivative of $\ln(K)$ with respect to $1/T$ for the VFT model obtained by peak fitting in the present work, the Arrhenius model obtained by numerical fitting of the peaks and via the Kissinger method in ref. [12]. The non-zero curvature of the VFT model is clearly shown.